\documentclass[conference]{IEEEtran}
\IEEEoverridecommandlockouts
\usepackage{ragged2e}
\usepackage{booktabs} 
\usepackage{soul}
\usepackage{threeparttable}
\usepackage{enumitem}
\usepackage{indentfirst}
\usepackage{multirow}
\usepackage{makecell}
\usepackage{balance}
\usepackage{cite}
\usepackage{tabularx}
\usepackage{caption}
\usepackage{amsmath,amssymb,amsfonts}
\usepackage{algorithm}
\usepackage{algpseudocode}
\usepackage{tikz}
\usepackage{newunicodechar}
\usepackage{tikz}
\usepackage{newunicodechar}
\usepackage{booktabs}
\usepackage{pifont}
\usepackage{collcell}
\usepackage{ifthen}
\newcommand{\cmark}{\ding{51}} 
\newcommand{\xmark}{--} 
\newcommand{\YN}[1]{\ifthenelse{\equal{#1}{Y}}{\cmark}{\xmark}}
\newcommand{\YH}[1]{\textcolor{black}{#1}}
\usepackage[T1]{fontenc}
\usepackage{newtxtext,newtxmath} 
\usepackage[hidelinks]{hyperref}

\newcommand{\blackcircled}[1]{%
  \tikz[baseline=(char.base)]{
    \node[shape=circle, fill=black, text=white, inner sep=0.4pt] (char) {\sffamily\bfseries\small #1};
  }%
}

\usepackage{eso-pic}
\usepackage{xcolor}

\newcommand{\preprintnotice}{%
  \AddToShipoutPictureFG{%
    \AtPageUpperLeft{%
      \raisebox{-0.75cm}[0pt][0pt]{%
        \makebox[\paperwidth][c]{\normalsize\color{gray} PREPRINT -- To appear at FPL 2026}%
      }%
    }%
  }%
}

\newunicodechar{❶}{\blackcircled{1}}
\newunicodechar{❷}{\blackcircled{2}}
\newunicodechar{❸}{\blackcircled{3}}
\newunicodechar{❹}{\blackcircled{4}}
\newunicodechar{❺}{\blackcircled{5}}
\newunicodechar{❻}{\blackcircled{6}}

\usepackage{graphicx}
\usepackage{textcomp}
\usepackage{xcolor}
\usepackage{booktabs}
\usepackage{array}      
\usepackage{makecell}   

\def\BibTeX{{\rm B\kern-.05em{\sc i\kern-.025em b}\kern-.08em
    T\kern-.1667em\lower.7ex\hbox{E}\kern-.125emX}}
\begin{document}

\title{

ExSpike: A General Full-Event Neuromorphic Architecture for Exploiting Irregular Sparsity with Event Compression
}
\author{
    Yuehai Chen and Farhad Merchant\\
    Bernoulli Institute and CogniGron, University of Groningen, The Netherlands\\
    Email: \{yuehai.chen, f.a.merchant\}@rug.nl
}
\preprintnotice
\maketitle
\begin{abstract}

Spiking neural networks (SNNs) promise energy-efficient computing due to their sparse spatio-temporal activity. However, effectively translating such irregular sparsity into practical performance and energy gains remains challenging, as full-event computing architectures are still underexplored. 
This paper proposes ExSpike, a general full-event neuromorphic architecture that fully exploits irregular sparsity in SNNs. To realize pure event-driven execution, we first propose a set of dataflow optimizations to ensure that the inputs to each SNN layer remain spike-based, thereby enabling full-event execution throughout the network. 
We then design a hardware-efficient full-event architecture, named ExSpike, which supports the optimized pure event-driven dataflow and an additional Attention Core for spike-driven self-attention. To further improve computing efficiency, we introduce adjacent-position event compression to reduce redundant accumulations across spatially adjacent spike sequences. ExSpike is implemented on an AMD Xilinx Virtex-7 FPGA and evaluated on both classification and segmentation workloads. Experimental results show that ExSpike achieves high normalized energy efficiency across diverse SNN models while maintaining competitive accuracy, delivering up to 479.15 GOPS, 281.85 GOPS/W, and 0.80 GOPS/W/PE. In particular, ExSpike achieves up to 10$\times$ higher PE-normalized energy efficiency than the SOTA FPGA-based SNN accelerator (FireFly-T). \YH{The code for ExSpike is available at \url{https://github.com/xiaoyuehai/ExSpike}}.

\end{abstract}


\section{Introduction}
Spiking neural networks (SNNs) were proposed in the 1990s and are regarded as the third generation of neural networks \cite{snn_algorithm2hardware}. 
In recent years, SNNs have gained significant attention because of their closer alignment with biological mechanisms and an efficient event-driven computation model. 
Recent advances in training algorithms and applications \cite{Zheng_Wu_Deng_Hu_Li_2021, knowledge_disll_snn_al, YE2025107790, zhao2025brillmbraininspiredlargelanguage} have significantly improved the capability of SNNs. 
In particular, recent object detectors based on spiking convolutional neural networks (SCNNs) have achieved 67.2\% mAP@50 on Gen1~\cite{detournemire2020large}, a dynamic vision sensor (DVS) autonomous driving dataset. This result is 2.5\% higher than that of an artificial neural network (ANN) with an equivalent architecture, while improving energy efficiency by 5.7$\times$.
 Furthermore, the development of spike-driven self-attention (SDSA) has enabled spiking transformers \cite{yao2023spike, zhou2026spikingformer, lee2025spiking} to emerge as a promising architecture for complex workloads. These advances, together with the inherent sparsity of SNNs, provide substantial opportunities for improving computational efficiency. However, modern general-purpose processors such as central processing units (CPUs) and graphics processing units (GPUs) are not well suited to the sparse, irregular computation and memory access patterns of SNNs, and thus cannot fully exploit their event-driven nature. This mismatch has motivated extensive research on specialized SNN hardware, including digital, analog, and mixed-signal neuromorphic processors \cite{3d_array, mint_aspdac, phi_isca}.
\begin{figure}[t] 
\centering
\includegraphics[width=\columnwidth]{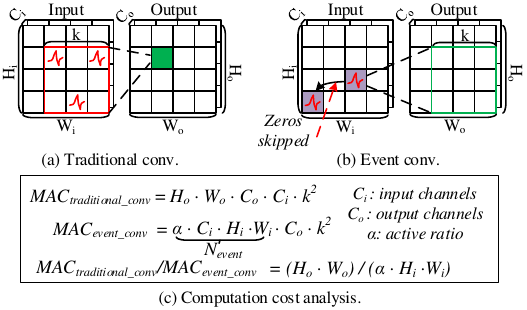} 
\caption{The computation model of different convolutions.}
\label{fig:comparison_tc_ec}
\end{figure}
As shown in Fig.~\ref{fig:comparison_tc_ec}, the dataflow in SCNNs can be divided into two fundamental types that differ in how convolution is triggered and propagated. Fig.~\ref{fig:comparison_tc_ec}(a) shows the traditional convolution (TConv) dataflow, in which each output neuron corresponds to a receptive field and requires weighted accumulation over all input channels within that region. In contrast, Fig.~\ref{fig:comparison_tc_ec}(b) illustrates the event-driven convolution (EConv) dataflow, where each input spike event has a fixed spatial influence range and simultaneously affects neurons at the same spatial location across all output channels. As a result, every operation directly contributes to valid updates, avoiding the workload imbalance introduced by neuron-level computation.

\begin{figure}[t] 
\centering
\includegraphics[width=\columnwidth]{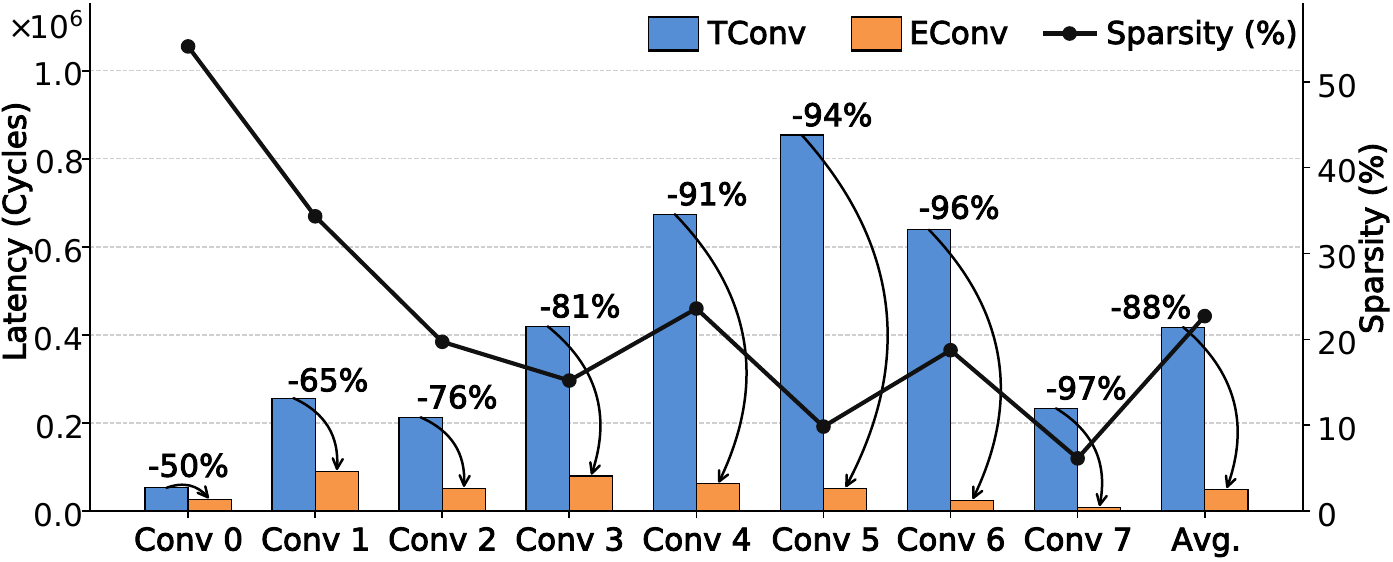} 
\caption{\YH{Layer-wise latency and sparsity analysis of VGG11 on CIFAR-10 under direct-coded SNN inference. Static CIFAR-10 images are encoded by the first layer into spike events.}}
\label{fig:tconv_vs_econv}
\end{figure}

Based on this distinction, Fig.~\ref{fig:comparison_tc_ec}(c) compares the computation cost of the two schemes. TConv incurs a fixed cost determined by the spatial dimensions and channel counts, whereas the cost of EConv scales with the number of active events and therefore benefits significantly from input sparsity. Fig.~\ref{fig:tconv_vs_econv} further reports the layer-wise latency of TConv and EConv in VGG11 \cite{simonyan2014very}, along with the corresponding input spike sparsity. EConv outperforms TConv in every layer, achieving up to 97\% latency reduction at Conv-7 and 88\% reduction on average. The results also show that higher input sparsity leads to larger speedups. For example, when only 10\% of spikes are active, corresponding to 90\% sparsity, the runtime can ideally be reduced by 90\%. Therefore, given the inherently low activity of SNNs, EConv can substantially reduce redundant accumulations and memory traffic while preserving correctness, thereby improving throughput and reducing energy consumption.

Although several neuromorphic processors for SNNs support sparsity-aware processing and direct-encoding acceleration~\cite{10521899,10992893,firefly-s}, three important challenges remain. 
\YH{First, direct coding is a widely used mechanism in SNNs, where the first layer acts as an encoder that receives multi-bit values rather than 1-bit spikes.} This requires multi-bit multiply-accumulate (MAC) operations. \YH{In addition, average pooling generates non-binary intermediate results because it divides the number of valid spikes within a pooling window by the pooling-window size.} Both operations interrupt pure event-driven execution.
Second, traditional convolution mapping is often inefficient and prone to workload imbalance, since each processing element (PE) is typically assigned to one output neuron; under irregular sparsity, this results in low PE utilization and increased latency. 
Third, full-event neuromorphic architectures are still underexplored, limiting the ability of existing designs to fully exploit the inherent sparsity of SNNs. \YH{In this work, full-event execution refers to an execution mode in which the main computations in SNNs, such as convolution and fully connected computation, are triggered by valid spike events and performed in an event-driven manner.
}

To address these challenges, we propose ExSpike, a general full-event neuromorphic architecture. By jointly optimizing dataflow and architecture, ExSpike reduces these bottlenecks and improves both energy efficiency and end-to-end performance. The main contributions are as follows.
\begin{enumerate}[leftmargin=*]
\item \textbf{\textit{Dataflow: }} We identify the key operations in SNN models that interrupt pure event-driven execution, and propose a set of optimizations to restore a fully event-driven dataflow, including direct coding, event-driven convolution, and event-driven average-pooling/fully connected computation. 

\item \textbf{\textit{Architecture: }} We propose ExSpike, a hardware-efficient full-event neuromorphic architecture that matches the optimized pure event-driven dataflow and integrates an Attention Core to support SDSA computation. In addition, we introduce adjacent-position event compression (APEC) to merge redundant events across adjacent spatial positions, thereby further reducing computation cost.

\item \textbf{\textit{Implementation: }} We implement ExSpike on an AMD Xilinx FPGA and evaluate it on diverse SNN workloads, including VGG11, ResNet18, spiking transformer models, and image segmentation networks. Experimental results show that ExSpike achieves up to 479.15 GOPS, 281.85 GOPS/W, and 0.80 GOPS/W/PE, while maintaining competitive accuracy across the evaluated workloads. These results demonstrate the efficiency of ExSpike and its support for diverse SNN models and operators.



\end{enumerate} 

\section{Pure Event-Driven Dataflow Optimization}
\label{opt_dataflow}
\YH{ExSpike uses leaky integrate-and-fire (LIF) neurons as the spiking neuron model.} Although the basic data representation in SNNs is spike-based, some operators or processing stages may generate non-spike intermediate results, which are then forwarded to subsequent layers and break the event-driven execution flow. Typical examples include direct coding and average pooling. In this section, we introduce a set of optimizations that enable pure event-driven execution of SNNs.

\subsection{Pure Event-Driven Convolutional Computing Dataflow}
\noindent\textbf{Optimization I: Direct Coding.}
The direct coding layer at the network input poses a major challenge to pure event-driven execution, because its input activations are multi-bit fixed-point values rather than binary spikes. For this reason, some previous work executed the coding layer on the host side and transferred only the generated spike sequences to the accelerator \cite{stisa, fireflyt, spikehards}. To avoid such off-chip preprocessing, we exploit the shift-and-accumulate formulation of binary multiplication to support direct coding within the accelerator. Specifically, as summarized under \textbf{Optimization I (OPT1)} in Algorithm~\ref{alg:pure_full_event}, the input activations are quantized into signed fixed-point values and then bit-sliced, while the weights are duplicated and shifted accordingly to match the bitwise multiplication process in the coding layer.

\noindent\textbf{Optimization II: Event-Driven Convolution.}
Unlike conventional convolution accelerators, where each PE is typically mapped to a neuron for kernel computation and then time-multiplexed to support all output neurons, the irregular sparsity of different kernels can lead to severe workload imbalance across PEs, while time multiplexing further increases latency. In contrast, in event-driven convolution, each valid spike event corresponds to a well-defined target update region. By partitioning the PE array into multiple blocks, each block can be dedicated to computing one target kernel region, so that every active cycle contributes directly to neural inference. The proposed event-driven convolutional dataflow is summarized under \textbf{Optimization II (OPT2)} in Algorithm~\ref{alg:pure_full_event}. For each convolutional layer, spike events are first collected at the same spatial location across input channels (line 9). If valid events exist, weight accumulation is performed by $\textsc{Cal}(e)$ according to their positions (lines 10--11). Furthermore, since a single spike event affects the same receptive-field region across all output channels, channel-level parallelism can be exploited to reduce latency while maintaining high PE utilization. When the number of output channels exceeds the available hardware parallelism, the architecture is reused across multiple groups to complete the updates for all output neurons (lines 5--6).

\subsection{Pure Event-Driven Fully Connected Computing Dataflow}

\noindent\textbf{Optimization III: Event-Driven Average-Pooling and Fully Connected Computation.}\label{OPT3}
In some SNN models, an average-pooling layer is inserted between the last convolutional layer and the first fully connected (FC) layer to preserve globally aggregated features and reduce the computational complexity of the FC layer. \YH{However, average pooling makes end-to-end event-driven computation difficult because it produces non-binary intermediate activations rather than spike events.} NEURAL \cite{11420315} proposed a window-to-time-to-first-spike (W2TTFS) mechanism to avoid the non-binary intermediate representation introduced by average pooling. However, this approach is not fully event-driven, since it still requires additional counters to record the number of valid events. In this work, we propose an event-driven average-pooling fully connected (EAFC) dataflow. As summarized under \textbf{Optimization III (OPT3)} in Algorithm~\ref{alg:pure_full_event}, we first scale the FC weight matrix according to the pooling size. Then, similar to event-driven convolution, we check the valid events at each spatial position and perform event-driven updates of the FC neurons $\mathbf{y}_{fc}$.

With these three optimizations, the SNN model can be executed in a pure event-driven manner without using multipliers. In the next section, we present the proposed full-event hardware architecture that supports these optimizations.

\begin{algorithm}[t]
\caption{Optimizations for Pure Full-Event Computation}
\label{alg:pure_full_event}
\footnotesize
\begin{algorithmic}[1]
\Require Input map $\mathbf{S}$ of an SNN layer; data precision $B$; convolution weight $\mathbf{W}$; FC weight $\mathbf{W}_{fc}$; output channels $C_o$; input size $H_i \times W_i$; architecture parallelism $P$
\Statex \textbf{Optimization I: Direct Coding}
\If{first\_layer}
    \State $\mathbf{S} \gets \mathbf{S}.\textsc{Quantize}(B).\textsc{BitSlice}(B)$
    \State $\mathbf{W} \gets \mathbf{W}.\textsc{DuplicateShift}(B)$
\EndIf

\Statex \textbf{Optimization II: Event-Driven Convolution}
\State $G \gets \lceil C_o / P \rceil$
\For{$g = 1$ to $G$}
    \For{$h = 1$ to $H_i$}
        \For{$w = 1$ to $W_i$}
            \For{\textbf{each} valid event $e$ in $\mathbf{S}[:,h,w]$}
                \State $\mathbf{psum}_g \gets \mathbf{W}_g.\textsc{Cal}(e)$ \Comment{active event}
                \State $\mathbf{MP}_g \gets \mathbf{MP}_g + \mathbf{psum}_g$
            \EndFor
        \EndFor
        \State $\mathbf{MP}_g.\textsc{CheckAndFire}()$ \Comment{output spikes}
    \EndFor
\EndFor

\Statex \textbf{Optimization III: Event-Driven Average-Pooling and Fully Connected Computation}
\State $\mathbf{W}_{fc} \gets \mathbf{W}_{fc} / pooling\_size^2$
\For{$h = 1$ to $H_i$}
    \For{$w = 1$ to $W_i$}
        \For{\textbf{each} valid event $e$ in $\mathbf{S}[:,h,w]$}
            \State $\mathbf{y}_{fc} \gets \mathbf{y}_{fc} + \mathbf{W}_{fc}.\textsc{Cal}(e)$
        \EndFor
    \EndFor
\EndFor
\end{algorithmic}
\end{algorithm}

\section{Architecture Design and Optimization}
\begin{figure}[t] 
\centering
\includegraphics[width=\columnwidth]{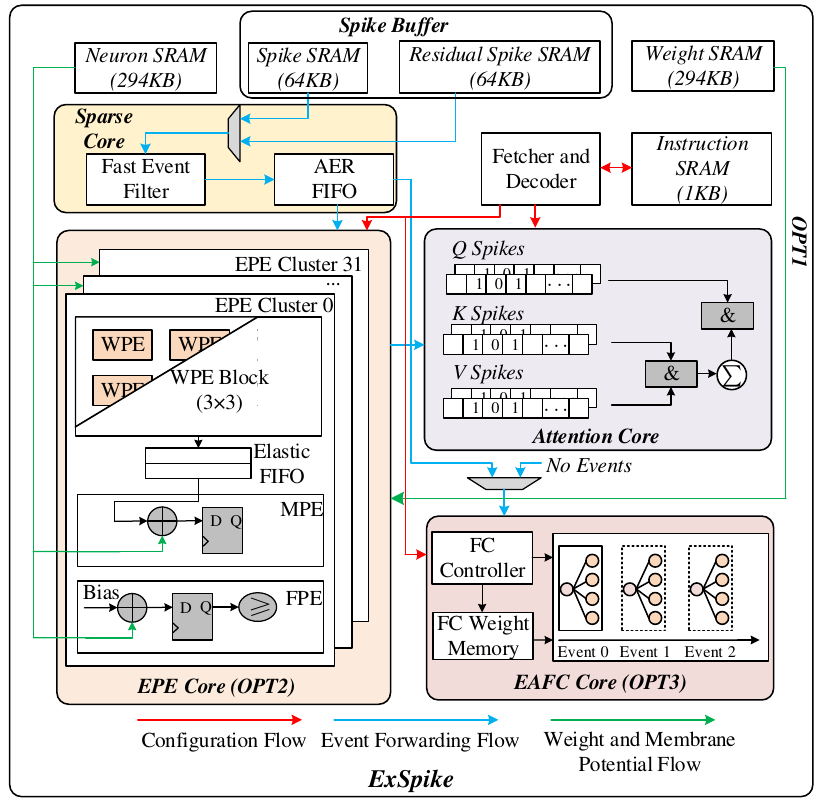} 
\caption{The overall architecture of ExSpike.}
\vspace{-10pt}
\label{fig:overall_arch}
\end{figure}

Fig.~\ref{fig:overall_arch} illustrates the overall architecture of ExSpike, which consists of four computing cores: the Sparse Core, the event-driven processing element (EPE) Core, the Attention Core, and the event-driven average-pooling and fully connected (EAFC) Core. \YH{In the Sparse Core, spike sequences are fetched from either the Spike SRAM or the Residual Spike SRAM, depending on the input source of the current SNN layer. The Residual Spike SRAM stores source spike feature maps used by shortcut or residual connections.} Valid event addresses are then extracted by the fast event filter and stored in the address-event representation (AER) FIFO. When the AER FIFO is non-empty, it triggers the computation of the EPE Core.

The EPE Core serves as the primary execution engine to support the optimized dataflow \textbf{\textit{(OPT2)}} and comprises 32 EPE clusters. Each EPE cluster contains a weight processing element (WPE) block, a membrane potential processing element (MPE), and a fire processing element (FPE). The WPE block performs weight accumulation and writes the results to the elastic FIFO (eFIFO). When the eFIFO is non-empty, the MPE updates the membrane potentials of neurons. The FPE then performs bias accumulation and threshold comparison to generate output spikes. In this way, each EPE cluster updates the membrane potential of one output channel, allowing the EPE Core to process 32 output channels in parallel and thereby improve computational throughput. 
\YH{Moreover, the EPE Core also supports \textbf{\textit{OPT1}}, where the duplicated and shifted weights are preprocessed offline and stored in the Weight SRAM. During inference, the EPE Core directly reads these preprocessed weights without additional logic.}

The Attention Core is responsible for spike-driven self-attention computation. It receives output spikes from the EPE Core and performs attention computation when the attention mechanism is enabled; otherwise, it directly writes the spikes back to the Spike Buffer. The EAFC Core supports the fused computation of average pooling and fully connected operations \textit{\textbf{(OPT3)}}, thereby enabling a full-event computing flow.

Finally, ExSpike can flexibly support different SNN topologies by programming the Instruction SRAM with model parameters, such as kernel size and channel count. The fetcher and decoder then read and decode the instructions to perform layer-by-layer inference.

\subsection{Event-driven Convolution Hardware Architecture}
\begin{figure}[t]
    \centering
    \includegraphics[width=\columnwidth]{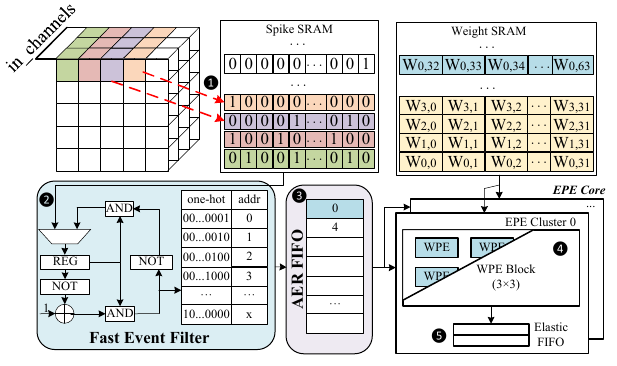}
    \caption{Hardware implementation of event-driven convolution.}
    \label{fig:event_computation}
\end{figure}
\subsubsection{Proposed Architecture Features}
Fig.~\ref{fig:event_computation} illustrates the hardware implementation of the proposed event-driven convolution dataflow, which consists of five main components. First, the spike sequence storage strategy ❶ is adopted, in which each address in the Spike SRAM stores spike data from all input channels at the same spatial location. Next, the fast event filter ❷ identifies one valid event position per cycle. Specifically, it first extracts the one-hot code corresponding to the lowest active bit and then uses this code as an index to obtain the corresponding valid position from a look-up table. The resulting event position is then written into the AER FIFO ❸. When the AER FIFO is non-empty, indicating the presence of valid events, the EPE Core fetches an event index from the FIFO, reads the corresponding weight matrix from the Weight SRAM, and performs weight accumulation. In the proposed architecture, the EPE Core contains 32 parallel EPE Clusters, enabling the accumulation of convolution weights for 32 output channels associated with the same target region in parallel. Accordingly, the Weight SRAM stores the convolution weights of different input channels across these 32 output channels. Each WPE block further integrates $3 \times 3$ WPE units to compute the weights of a single convolution kernel ❹. Once the AER FIFO becomes empty and the fast event filter is idle, both event filtering and event computation are complete. At this point, the partial sums generated by the EPE clusters are written into the elastic FIFO ❺, which serves as the input source for the subsequent MPE stage.
\begin{figure}[t] 
\centering
\includegraphics[width=\columnwidth]{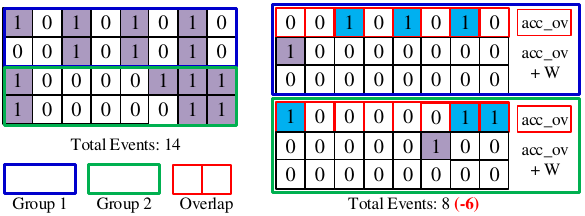} 
\caption{Implementation of adjacent-position event compression.}
\label{fig:Event_Compression}
\end{figure}

\subsubsection{Adjacent-position Event Compression (APEC)}
\label{APEC_theory}

To further improve event computing efficiency, we propose a novel event compression mechanism, namely adjacent-position event compression (APEC). Input spike sequences are grouped by spatial adjacency. As illustrated in Fig.~\ref{fig:Event_Compression}, adjacent spatial positions are partitioned into groups, and each group is compressed by extracting the shared overlap among its spike sequences. For each group, we first apply bit-wise AND logic to identify the overlap sequence, and then derive the corresponding non-overlap sequences that are disjoint from the overlap. During computation, the overlap sequence is processed first and its partial sums are cached, after which the non-overlap sequences are updated using only their unique event contributions. In this way, repeated accumulations caused by overlapped events can be avoided, thereby reducing both memory traffic and computation cost.

Fig.~\ref{fig:Event_Compression} provides a concrete example. After compression, the total number of events is reduced from 14 to 8. For a $3\times3$ convolution with 64 output channels, this directly eliminates $6\times64\times9=3456$ accumulation operations. Since APEC only reorganizes the execution order of overlapped and non-overlapped events, it preserves numerical equivalence with the original full-event execution. For theoretical analysis, we consider a single APEC group of size $g$. APEC is motivated by the observation that adjacent spatial positions often exhibit correlated spike activities, which leads to a non-negligible overlap among their spike sequences. Let $S_i \subseteq \{1,2,\dots,C_i\}$ denote the active-channel set of the $i$-th spike sequence, where $C_i$ is the number of input channels. The common overlap of the group is defined as
\begin{equation}
O_G=\bigcap_{i=1}^{g} S_i.
\end{equation}
Let $\Phi(c)$ denote the convolution contribution of an input event $c$. Since each spike sequence can be decomposed into the shared overlap and its non-overlap part, APEC computes the overlap only once and then adds the remaining unique contributions of each sequence individually. Therefore, APEC is numerically equivalent to the original full-event execution.

Accordingly, the number of eliminated redundant events is
\begin{equation}
\Delta N_{\mathrm{event}}^{(g)}=(g-1)|O_G|.
\end{equation}
Since each valid input spike event affects all output channels over all $k^2$ kernel positions, the corresponding accumulation reduction brought by APEC can be expressed as
\begin{equation}
\Delta C^{(g)}=\Delta N_{\mathrm{event}}^{(g)}C_ok^2=(g-1)|O_G|C_ok^2.
\end{equation}
This equation shows that the gain of APEC is jointly determined by the group size and the overlap size, where $|O_G|$ directly reflects the correlation among adjacent spike sequences.

The main overhead of APEC lies in the storage and reuse of overlap partial sums, which can be approximated as
\begin{equation}
M_{\mathrm{ov}} \approx C_ok^2w_{\mathrm{acc}},
\end{equation}
where $w_{\mathrm{acc}}$ denotes the bitwidth of the partial sum. Therefore, although a larger group provides a higher theoretical reuse factor, the overall gain of APEC does not necessarily increase monotonically with group size because the higher-order overlap $|O_G|$ typically decreases as more spike sequences are included. This implies the existence of an optimal group size, as discussed in Section~\ref{Event_evaluation}.

\subsection{Implementation of EAFC Core}
As shown in Fig.~\ref{fig:overall_arch}, when an average-pooling layer is followed by a fully connected (FC) layer, the EAFC Core is activated to process valid events forwarded from the Sparse Core. Specifically, upon receiving a valid event, the FC Controller generates the corresponding read address for the FC Weight Memory and triggers FC computation. Therefore, the FC computation is carried out in a fully event-driven manner. To fuse average pooling with the FC layer, the FC weights are scaled offline during the generation of ExSpike configuration files. For example, for a pooling window of size $4 \times 4$, each FC weight is divided by 16 before the FC weight configuration files are generated.

\subsection{Implementation of Attention Core}
\begin{figure}[t] 
\centering
\includegraphics[width=\columnwidth]{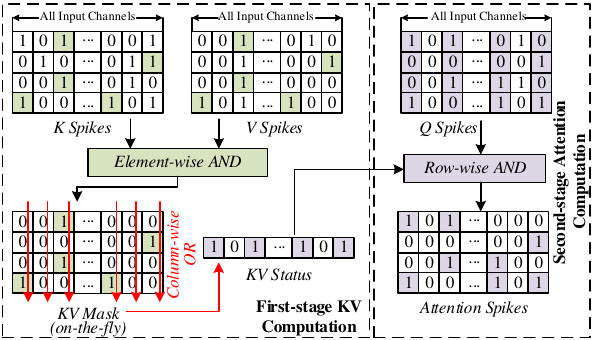} 
\caption{Implementation of the Attention Core.}
\label{fig:attention_core}
\end{figure}

Fig.~\ref{fig:attention_core} illustrates the hardware implementation of the Attention Core. Since attention-spike generation involves the processing of Q, K, and V spike sequences, the computation is divided into two stages. In the first stage, ExSpike performs KV computation. The K spikes are generated by FPEs first, followed by the V spikes. An element-wise AND operation between the K and V spikes is then used to produce the KV mask, and a column-wise OR operation is further applied to generate the KV status vector. \YH{The KV status vector is implemented using local registers rather than BRAM, because its size is small and only scales with the number of output channels, i.e., \(C_o\) bits.}
Importantly, no large intermediate buffer is required for storing the internal spike sequences, because these logic operations can be performed during the write-back of the V spikes while reading the corresponding K spikes. For example, when one row of V spikes is written back, the corresponding row of K spikes can be read out simultaneously, and the AND and OR operations are performed on the fly to update the KV status vector. In the second stage, we compute the attention spikes by performing the AND operation between each row of the Q spikes and the shared KV status vector.

\begin{figure*}[t] 
\centering
\includegraphics[width=\textwidth]{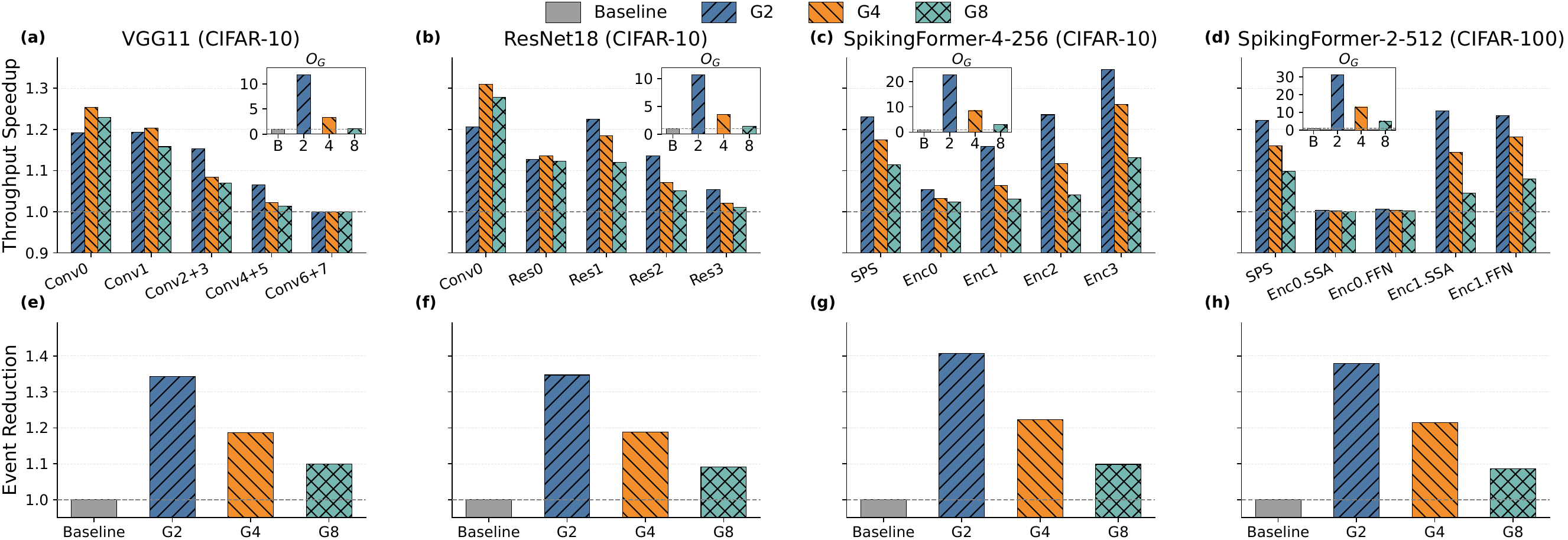} 
\caption{Performance analysis under different APEC group sizes. (a)--(d) show the block-wise throughput speedup over the baseline without APEC for VGG11, ResNet18, SpikingFormer-4-256, and SpikingFormer-2-512, respectively. VGG11 is divided into five convolutional blocks (Conv$x$), ResNet18 is divided by residual blocks (Res$x$), SpikingFormer-4-256 is divided into the spiking patch splitting (SPS) stage and encoder blocks (Enc$x$), and SpikingFormer-2-512 is partitioned more finely into SPS, spike-driven self-attention (SSA), and feedforward network (FFN) components. (e)--(h) show the corresponding total event reduction compared with the baseline without APEC.} 
\label{fig:compare_to_sibrain}
\end{figure*}
\section{Experimental Results}
\label{sec:experiments}
We implemented and evaluated four spiking models for classification, including VGG11 on CIFAR-10, SpikingFormer-4-256 on CIFAR-10, ResNet18 on CIFAR-10, and SpikingFormer-2-512 on CIFAR-100. For segmentation, we designed a spiking SegNet \cite{9855834} with the architecture 8C3--16C3--32C3--32C3--16TC3--2TC3, where $XCY$ denotes a convolution layer with $X$ output channels and a kernel size of $Y \times Y$, and $TC$ denotes transposed convolution. All models were trained using AdamW with a batch size of 128. The training was conducted in PyTorch and SpikingJelly \cite{doi:10.1126/sciadv.adi1480} using LIF neurons ($\tau = 0.5$) on 4 NVIDIA RTX PRO
6000 GPUs.
ExSpike was designed in Verilog HDL, synthesized using Synplify, and implemented in Xilinx Vivado. \YH{The weight precision is 8-bit fixed-point and the membrane potential precision is 16-bit fixed-point. For power estimation, we used the post-synthesis netlist and benchmark-specific SAIF files generated from simulation. The power consumption was then estimated using Vivado.} The final design runs at 200\,MHz on an AMD Xilinx Virtex-7 XC7V2000T FPGA. 

\subsection{Efficiency of Event Execution with Compression}
\label{Event_evaluation}

Fig.~\ref{fig:compare_to_sibrain} evaluates the efficiency of APEC under different group sizes in the VGG11, ResNet18, and SpikingFormer models. In general, APEC consistently improves throughput on all benchmarks, while GROUP=2 (G2) achieves the best result in every case. Compared to baseline, G2 improves average throughput by 10.9\%, 13.8\%, 14.5\%, and 11.2\% on VGG11 (CIFAR-10), ResNet18 (CIFAR-10), SpikingFormer-4-256 (CIFAR-10), and SpikingFormer-2-512 (CIFAR-100), respectively. Meanwhile, the corresponding event reduction ratios are 1.62$\times$, 1.35$\times$, 1.36$\times$, and 1.38$\times$, confirming that APEC effectively reduces the number of valid events to be executed and translates this reduction into throughput gains.

The inset plots further reveal that the average $O_G$ decreases rapidly as the group size increases. For example, on SpikingFormer-4-256 (CIFAR-10), the average $O_G$ decreases from 19.08 for G2 to 6.82 for GROUP=4 (G4) and 2.92 for GROUP=8 (G8). A similar trend is observed on all other benchmarks. This directly validates the theoretical analysis in Section~\ref{APEC_theory}: although a larger group provides a higher theoretical reuse factor, the higher-order overlap shrinks quickly as more adjacent spike sequences are grouped together, which weakens the practical compression gain. As a result, the overall benefit of APEC does not increase monotonically with the group size, and G2 provides the best practical trade-off between the reuse factor and the rapidly diminishing higher-order overlap. Based on these results, G2 is adopted as the default APEC configuration in ExSpike.

\begin{figure}[t] 
\centering
\includegraphics[width=\columnwidth]{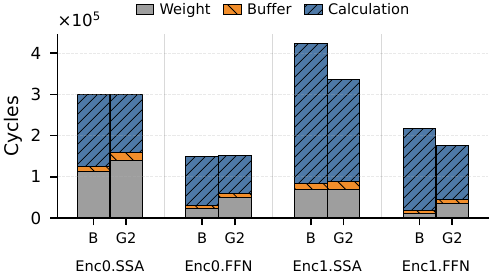} 
\caption{Latency breakdown for SpikingFormer-2-512 under Baseline and APEC-2. For each block, the total execution latency is decomposed into waiting-for-weight-ready (Weight) cycles, buffer cycles, and calculation cycles.}
\label{fig:latency_breakdown}
\vspace{-12pt}
\end{figure}

Another notable observation is that event reduction does not always translate into proportional throughput improvement for every layer. Some layers, such as Conv6+7 in VGG11 and Enc0 in SpikingFormer, exhibit reduced event counts but only marginal throughput changes, suggesting that their performance is constrained by other architectural bottlenecks beyond event-driven accumulation alone. As a concrete example, Fig.~\ref{fig:latency_breakdown} shows the cycle-level latency breakdown of SpikingFormer-2-512 before and after applying APEC-2. Although APEC-2 reduces the calculation cycles in both Enc0.SSA and Enc0.FFN, the Weight cycles increase significantly, offsetting the computation reduction and resulting in limited end-to-end latency improvement. Therefore, 
APEC-G2 is beneficial when the saved calculation cycles exceed the additional weight/buffer cycles; otherwise, the increased weight-ready latency may offset the event-compression gain, making APEC-G2 more suitable for computation-bound layers with strong adjacent-position spike overlap.
\begin{table}[t]
\centering
\footnotesize
\setlength{\tabcolsep}{2.4pt}
\renewcommand{\arraystretch}{1}
\caption{Resource breakdown of ExSpike. Values are reported as Baseline/APEC-2 where applicable. Power reports the total dynamic power of the entire ExSpike accelerator.}
\label{tab:resource_breakdown}
\begin{tabular}{lccccc}
\toprule
 & EPE Core & Attention Core & EAFC Core & Sparse Core & Total \\
\midrule
kLUTs     & 19 / 25 & 6 & 3   & 4     & 36 / 41 \\
kFFs      & 21 / 26 & 4 & 1   & 2 / 3 & 36 / 41 \\
BRAMs     & 128   & 0 & 2.5 & 0 & 312   \\
\bottomrule
Power (W) & \multicolumn{5}{c}{1.593 / 1.700} \\
\bottomrule
\end{tabular}
\end{table}

\begin{table*}[!t]
\centering
\footnotesize
\setlength{\tabcolsep}{3pt}
\renewcommand{\arraystretch}{0.8}
\caption{Comparison with state-of-the-art FPGA-based SNN accelerators.}
\label{tab:fpga_compare}

\resizebox{\textwidth}{!}{%
\begin{tabular}{llc ccc ccccc ccccc}
\toprule
\multirow{2}{*}{Work} & \multirow{2}{*}{Device} & \multirow{2}{*}{Clk (MHz)} 
& \multicolumn{3}{c}{Benchmark} 
& \multicolumn{5}{c}{Resource$^1$} 
& \multicolumn{5}{c}{Performance$^2$} \\
\cmidrule(lr){4-6} \cmidrule(lr){7-11} \cmidrule{12-16}
& & 
& Dataset & Model & Acc. 
& PE Size & kLUTs & kFFs & BRAM & DSPs 
& FPS & GOPS & GOPS/W & \makecell[c]{GOPS/W\\/PE} & \makecell[c]{GOPS/W\\/kLUTs} \\ 
\midrule

\multirow{2}{*}{Cerebron \cite{9855834}} 
& \multirow{2}{*}{xc7z100} 
& \multirow{2}{*}{200} 
& CIFAR-10 & MobileNet & 91.90 
& 256 & 85 & 70 & 283 & 0 
& 90 & 44.2 & 31.6 & 0.12 & 0.37 \\
& & 
& MLND\_Capstone & SegNet$^3$ & 97.30 
& 256 & 85 & 70 & 283 & 0 
& 1250 & 45.0 & 32.1 & 0.13  & 0.38\\
\midrule

DeepFire2 \cite{deepfire} 
& xcvu9p 
& 550 
& CIFAR-10 & VGG10 & 87.10 
& -- & 125 & -- & 511 & 2025 
& 23255 & 10400 & 517.93 & -- & 0.44\\
\midrule

SpikeTA \cite{spikeTA}
& xcu280 
& 300 
& CIFAR-100 & SpikingFormer-2-512$^5$ & 78.40 
& 35622 & 503 & -- & 1366 & 7249 
& -- & 28980 & 405.60 & 0.01 & 0.10\\
\midrule

NEURAL \cite{11420315}
& xc7v2000t 
& 200 
& CIFAR-10 & VGG11 & 93.45 
& 256 & 74 & 63 & 137.5 & 0 
& 68 & 41.37 & 52.37 & 0.20 & 0.71 \\
\midrule

STISA \cite{stisa}
& xczu9eg 
& 200 
& CIFAR-10 & ResNet18 & 95.29 
& 387 & 137 & 137 & 160 & 0 
& 49 & 126.28 & 74.72 & 0.19 & 0.55\\
\midrule

\multirow{2}{*}{FireFly-T \cite{fireflyt}} 
& \multirow{2}{*}{xczu5ev} 
& \multirow{2}{*}{300} 
& CIFAR-10 & SpikingFormer-4-256$^5$ & 94.45 
& 8192 & 46 & 117 & 100 & 304 
& 907 & 3029 & 696.64 & 0.08 & 3.43\\
& & 
& CIFAR-100 & SpikingFormer-2-512$^5$ & 78.38 
& 8192 & 46 & 117 & 100 & 304 
& 630 & 3317 & 762.87 & 0.09 & 3.75\\
\midrule

\multirow{2}{*}{SConvNSys \cite{11247941}} 
& \multirow{2}{*}{xcu250} 
& \multirow{2}{*}{250} 
& CIFAR-10 & VGG11 & 91.48 
& 1024 & 308 & 243 & 324.5 & 0 
& 73 & 183.17 & 105.3 & 0.10 & 0.34\\
& & 
& MLND\_Capstone & USegNet$^4$ & 99.00 
& 800 & 312 & 257 & 287 & 3 
& 154 & 43.5 & 24.2 & 0.03 & 0.08\\
\midrule

\multirow{5}{*}{ExSpike} 
& \multirow{5}{*}{xc7v2000t} 
& \multirow{5}{*}{200} 
& CIFAR-10 & VGG11 & 93.89 
& 352 & 41 & 40 & 312 & 0 
& 148 & 361.51 & 211.91 & 0.60 & 5.17\\
& & 
& CIFAR-10 & ResNet18 & 94.98 
& 352 & 41 & 40 & 312 & 0 
& 85 & 209.31 & 120.64 & 0.34 & 2.94\\
& & 
& CIFAR-10 & SpikingFormer-4-256$^5$ & 94.45 
& 352 & 40 & 40 & 312 & 0 
& 197 & 479.15 & 281.85 & 0.80 & 7.05\\
& & 
& CIFAR-100 & SpikingFormer-2-512$^5$ & 75.81 
& 352 & 41 & 40 & 335 & 0 
& 51 & 463.90 & 267.53 & 0.76 & 6.53\\
& & 
& MLND\_Capstone & SegNet$^3$ & 98.70 
& 352 & 43 & 44 & 310 & 0 
& 1633 & 123.25 & 82.78 & 0.24 & 1.93\\
\bottomrule
\end{tabular}%
}

\begin{minipage}{\linewidth}
\raggedright\scriptsize
\justifying
\vspace{2pt}
\noindent
$^1$ PE size denotes the total number of processing elements. The reported kLUTs/kFFs/BRAM/DSPs denote the total resource usage of the entire architecture.
$^2$ FPS: frames per second. For a fair comparison, we report normalized efficiencies in GOPS/W/PE and GOPS/W/kLUTs. For GOPS/W/kLUTs, the kLUT count includes both the used LUT resources and the DSP-to-LUT equivalence. Following~\cite{dsp-to-luts}, one DSP is converted to 483 LUTs for Xilinx 7-series (25b$\times$18b) and 517 LUTs for Xilinx UltraScale+ (27b$\times$18b) devices.
$^3$ SegNet: 8C3--16C3--32C3--32C3--16TC3--2TC3. 
$^4$ USegNet: 16C3--16C3--64C3--MP2--128C3--MP2--64TC3--16TC3--1TC3.
$^5$ SpikingFormer-L-D means a spiking transformer model with L spiking transformer encoder blocks and D feature embedding dimensions.
\end{minipage}
\end{table*}

Table~\ref{tab:resource_breakdown} presents the detailed resource breakdown and total power consumption under two architectural configurations, including Baseline and APEC-2. As discussed in Section~\ref{APEC_theory}, the main overhead of APEC comes from the additional buffers used to store the partial sums of overlapped sequences. Therefore, when ExSpike is configured with an APEC group size of 2, the resource usage of the EPE Core increases accordingly, with LUT usage rising from 19k to 25k and FF usage increasing from 21k to 26k, while the resources of the other cores remain unchanged. This is because the EPE Core is the main computation engine that supports APEC execution. Meanwhile, the total power consumption increases by 1.07$\times$ due to the additional hardware overhead. However, this increase does not offset the performance gain provided by APEC. In particular, for SpikingFormer-4-256 on CIFAR-10, APEC improves throughput by 1.14$\times$, still resulting in an overall energy-efficiency improvement of 1.07$\times$.

\subsection{Comprehensive Evaluation with Prior Work}
\subsubsection{Performance and Energy Efficiency Analysis} 
Table~\ref{tab:fpga_compare} compares ExSpike with state-of-the-art FPGA-based SNN accelerators in terms of accuracy, throughput, and energy efficiency. For conventional SCNN models on CIFAR-10, ExSpike achieves the best overall balance between accuracy and efficiency. Specifically, for VGG11, ExSpike attains 93.89\% accuracy, improving upon NEURAL~\cite{11420315} and SConvNSys~\cite{11247941} by 0.44\% and 2.41\%, respectively. Meanwhile, it also delivers the highest throughput and energy efficiency among these VGG-style implementations, reaching 148 FPS and 211.91 GOPS/W.
For ResNet18 on CIFAR-10, STISA~\cite{stisa} achieves slightly higher accuracy by exploiting temporal optimization to reduce the effective timesteps of different network blocks. Since its reported result is based on a compressed timestep configuration, we set the inference timestep of ExSpike to be consistent with that setting for a fair comparison. Under this condition, ExSpike achieves 1.73$\times$ higher throughput and improves energy efficiency by 45.92 GOPS/W, while using a smaller PE size. Although its accuracy is 0.31\% lower than that of STISA~\cite{stisa}, ExSpike achieves substantially higher normalized energy efficiency in terms of GOPS/W/PE and GOPS/W/kLUTs, indicating better architectural efficiency under constrained resources.

For emerging spiking transformer models, ExSpike also demonstrates competitive deployment capability. On SpikingFormer-4-256 for CIFAR-10, ExSpike achieves 94.45\% accuracy, matching FireFly-T~\cite{fireflyt}. Although FireFly-T reports higher FPS and GOPS/W, its performance strongly relies on DSP-intensive optimization and a much larger PE array, which improves raw throughput but weakens portability to DSP-limited platforms and complicates ASIC migration. In contrast, ExSpike adopts a DSP-free full-event computing architecture and still achieves the highest normalized efficiency, reaching 0.80 GOPS/W/PE, which is about 10$\times$ higher than FireFly-T. A similar trend can be observed on SpikingFormer-2-512 for CIFAR-100. While ExSpike reports lower accuracy than SpikeTA~\cite{spikeTA} and FireFly-T~\cite{fireflyt}, it delivers much higher normalized energy efficiency per PE. Specifically, ExSpike achieves 0.76 GOPS/W/PE, which is nearly 76$\times$ and 8.4$\times$ higher than SpikeTA and FireFly-T, respectively. This result suggests that ExSpike is particularly effective in converting irregular event sparsity into hardware efficiency, even when deployed on more complex transformer-style SNNs.

For the segmentation workload, ExSpike also shows strong performance. On MLND\_Capstone with SegNet, ExSpike achieves 98.70\% accuracy, outperforming Cerebron~\cite{9855834} by 1.40\%, while also providing higher throughput and energy efficiency. In terms of normalized efficiency, ExSpike improves GOPS/W/PE by about 1.85$\times$, further demonstrating that the proposed architecture is effective not only for image classification but also for segmentation tasks. \YH{Across all tested workloads, ExSpike attains the highest GOPS/W/kLUTs, with values ranging from 1.93 to 7.05 GOPS/W/kLUTs. Its peak efficiency of 7.05 GOPS/W/kLUTs on SpikingFormer-4-256 surpasses FireFly-T~\cite{fireflyt} by 2.06$\times$.
}

\subsubsection{Resource Analysis}
As summarized in Table~\ref{tab:fpga_compare}, ExSpike achieves higher hardware efficiency with a compact implementation, requiring only about 41k LUTs and 40k FFs across multiple workloads while supporting diverse SNN models and applications. In contrast, SConvNSys~\cite{11247941} relies on multi-timestep parallel execution, which significantly increases logic cost to more than 300k LUTs and over 240k FFs, limiting its practicality on resource-constrained FPGA platforms. Cerebron~\cite{9855834} improves utilization through additional scheduling and control mechanisms, but this also introduces extra hardware overhead. FireFly-T~\cite{fireflyt} and SpikeTA~\cite{spikeTA} further depend on DSPs to boost raw throughput, which increases platform coupling and reduces portability under limited DSP budgets. By comparison, ExSpike adopts a DSP-free full-event computing architecture that performs online event detection and valid-event generation in one cycle, and completes SNN computation mainly through lightweight accumulators and efficient event-driven control. 

\subsubsection{Architecture Flexibility and Generality}
\YH{At the application level, ExSpike supports both image classification and segmentation, as demonstrated by its results on CIFAR-10/CIFAR-100 classification and MLND\_Capstone segmentation in Table II.} At the model level, ExSpike supports a broad range of mainstream SNN topologies, including fully connected networks, SCNNs, residual networks, and spiking transformer models. This flexibility is enabled by the programmable Instruction SRAM and the unified full-event computing architecture, which together support diverse operators such as fully connected, convolution/transposed convolution, shortcut connection, max/average pooling, and spike-driven self-attention. Therefore, compared with prior FPGA-based SNN accelerators that are often specialized for limited network structures or application scenarios, ExSpike provides a general and configurable neuromorphic acceleration platform, making it well-suited for exploring practical and diverse SNN deployments.

\begin{figure}[t] 
\centering
\includegraphics[width=0.9\columnwidth]{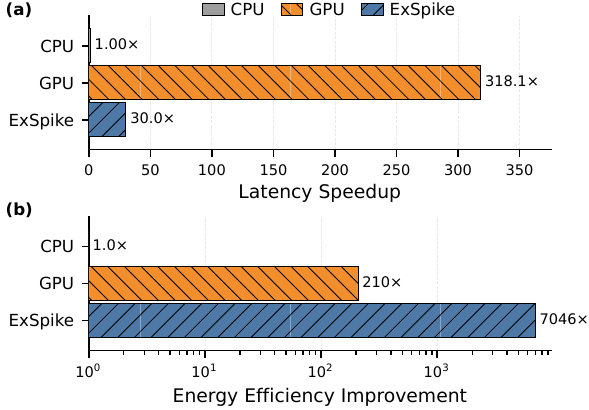} 
\caption{Performance comparison with Intel(R) Xeon 8470Q CPU and NVIDIA RTX PRO 6000 GPU.}
\label{fig:comparison_cpu_gpu}
\end{figure}
\vspace{-0.3em}
\subsection{Comparison with CPU and GPU}
Fig.~\ref{fig:comparison_cpu_gpu} compares the performance of ExSpike with an Intel(R) Xeon 8470Q CPU and an NVIDIA RTX PRO 6000 96GB GPU using the SpikingFormer-4-256 classification model on CIFAR-10. Compared with the CPU, ExSpike achieves 30.0$\times$ lower latency and 7046$\times$ higher energy efficiency, thanks to its specialized full-event architecture for SNNs. Compared with the GPU, ExSpike exhibits higher latency because the GPU executes the model with a larger batch size (32 in this comparison). Nevertheless, ExSpike still delivers nearly 33.6$\times$ higher energy efficiency, demonstrating its superior architectural efficiency for SNN inference.

\section{Related Work}
Event-driven neuromorphic architectures perform computation only after detecting valid spike events, ensuring efficient inference. The spiking multilayer perceptron model (SMLP) is highly suitable for event-driven computation \cite{9745256, 10981802, 10794606}, since the target region of a single spike is continuous and spans multiple neurons across layers, MP updates can be efficiently performed in parallel and pipelined for higher throughput. However, the representational capacity of SMLP remains limited and can only process simple input patterns. 

SCNNs address these limitations by enabling spatially structured, convolution-based spike processing \cite{10695026, 9785601, 9556367}. Some approaches \cite{10521899, 11420315, 9855834, 11247941, deepfire} improve performance by feeding valid spike sequences into computational engines through zero filtering and data reordering under TConv dataflows. Nevertheless, because these architectures are still primarily built on systolic-array- or adder-tree-based computing units \cite{deepfirev1_fpl, 11329523}, they do not fundamentally align execution with sparse spike events, and their processing units continue to suffer from workload imbalance and limited efficiency under irregular sparsity. 

Other approaches switch to EConv \cite{10992893, 10558062, 11043247}, typically by configuring dedicated hardware units for each layer. For direct-coding SCNNs, execution is often divided between separate dense and sparse cores \cite{10992893}, which reduces both resource utilization and computational efficiency. In addition, implementing EConv on general-purpose architectures struggles to maintain the low-power and high-throughput advantages of event-driven computation \cite{11043247, 10764529}. SpikeTA \cite{spikeTA} and FireFly--T \cite{fireflyt} are among the latest architectures delivering high computational performance for spiking transformer models; however, they rely heavily on DSP-based optimizations and do not fully exploit purely event-driven computation, which constrains their per-PE energy efficiency.

\begin{table}[t]
\centering
\scriptsize
\setlength{\tabcolsep}{2.6pt}
\renewcommand{\arraystretch}{1.10}
\caption{Feature comparison of representative FPGA-based SNN accelerators.}
\label{tab:related_work}
\begin{tabular}{lccccccc}
\toprule
Feature
& \makecell{NEURAL\\\cite{11420315}}
& \makecell{STISA\\\cite{stisa}}
& \makecell{DeepFire2\\\cite{deepfire}}
& \makecell{SpikeTA\\\cite{spikeTA}}
& \makecell{Cerebron\\\cite{9855834}}
& \makecell{FireFly--T\\\cite{fireflyt}}
& \makecell{\textbf{ExSpike}} \\
\midrule
Sparse-aware & \cmark & \xmark & \xmark & \xmark & \cmark & \cmark & \cmark \\
Event conv.  & \xmark & \xmark & \xmark & \xmark & \xmark & \xmark & \cmark \\
Full-event   & \xmark & \xmark & \xmark & \xmark & \xmark & \xmark & \cmark \\
SDSA         & \xmark & \xmark & \xmark & \cmark & \xmark & \cmark & \cmark \\
Flex. op.    & \xmark & \cmark & \cmark & \cmark & \cmark & \cmark & \cmark \\
Dir. coding  & \cmark & \xmark & \cmark & \xmark & \xmark & \xmark & \cmark \\
DSP-free     & \cmark & \cmark & \xmark & \xmark & \cmark & \xmark & \cmark \\
\bottomrule
\end{tabular}

\begin{minipage}{\linewidth}
\raggedright\scriptsize
\vspace{2pt}
Event conv.: Event-driven convolution; SDSA: Spike-driven self-attention computation; Flex. op.: Flexible operator support; Dir. coding: Direct coding support.
\end{minipage}
\end{table}

As shown in Table~\ref{tab:related_work}, compared with prior designs, ExSpike targets irregular sparse event dataflow and introduces event compression (EC) under the correctness guarantee of full-event execution, which significantly reduces ineffective accumulations and memory accesses, thereby improving throughput and energy efficiency. Beyond common spiking and convolution operators, ExSpike can support flexible operators such as transposed convolution and spike-driven self-attention computation. Overall, the architecture maintains generality while fully exploiting irregular sparsity to deliver efficient and scalable acceleration for diverse SNN workloads.

\section{Conclusion}
We proposed ExSpike, a full-event neuromorphic architecture developed through dataflow and architecture co-design. By optimizing the key operations that interrupt event-driven execution, ExSpike enabled a pure event-driven dataflow for SNN models. The proposed architecture supported diverse operators, including event-driven convolution and SDSA, while further reducing redundant valid events through APEC. Experimental results demonstrated that ExSpike effectively exploited irregular spike sparsity across diverse SNN workloads, including both image classification and segmentation models. ExSpike also achieved competitive performance with low hardware resource cost, while providing high architectural flexibility and generality for practical neuromorphic deployment.
\balance
\bibliographystyle{ieeetr}
\bibliography{references}

\end{document}